\documentclass[aps,prl,showpacs,twocolumn,superscriptaddress]{revtex4}
\usepackage{graphicx}
\usepackage{bm}

\def\be{\begin{equation}}
\def\ee{\end{equation}}
\def\bea{\begin{eqnarray}}
\def\eea{\end{eqnarray}}

\begin{document}
\title{Current-induced dynamics of spiral magnet}

\author{Kohei Goto}
\email{goto@appi.t.u-tokyo.ac.jp}
\affiliation{Department of Applied Physics, The University of Tokyo,
7-3-1, Hongo, Bunkyo-ku, Tokyo 113-8656, Japan}

\author{Hosho Katsura}
\email{katsura@appi.t.u-tokyo.ac.jp}
\affiliation{Department of Applied Physics, The University of Tokyo,
7-3-1, Hongo, Bunkyo-ku, Tokyo 113-8656, Japan}

\author{Naoto Nagaosa}
\email{nagaosa@appi.t.u-tokyo.ac.jp}
\affiliation{Department of Applied Physics, The University of Tokyo,
7-3-1, Hongo, Bunkyo-ku, Tokyo 113-8656, Japan}
\affiliation{Cross-Correlated Materials Research Group (CMRG), ASI,
RIKEN, Wako 351-0198, Japan}

\begin{abstract}
We study the dynamics of the spiral magnet under the charge current by solving 
the Landau-Lifshitz-Gilbert equation numerically. In the steady state, the current ${\vec j}$ induces
(i) the parallel shift of the spiral pattern with velocity $v=(\beta/\alpha )j$ 
($\alpha$, $\beta$: the Gilbert damping coefficients),
(ii) the uniform magnetization $M$ parallel or 
anti-parallel to the current depending on the chirality of the spiral and the 
ratio $\beta / \alpha $, and 
(iii) the change in the wavenumber $k$ of the spiral.
These are analyzed by the continuum effective theory using the scaling
argument, and the various nonequilibrium phenomena such as the chaotic behavior
and current-induced annealing are also discussed.
\end{abstract}

\maketitle
The current-induced dynamics of the magnetic structure is attracting intensive
interests from the viewpoint of the spintronics. 
A representative example is the current-driven motion of the magnetic
domain wall (DW) in ferromagnets \cite{Berger1,Berger2}. This phenomenon can be 
understood from the conservation of the spin angular momentum, i.e.,
spin torque transfer mechanism 
\cite{Slon,TataraKohno,Barnes,Yamaguchi,Yamanouchi}. 
The memory devices using this current-induced magnetic DW motion 
is now seriously considered \cite{Parkin}. 
Another example is the motion of the vortex structure on the disk of a ferromagnet, 
where the circulating motion of the vortex core is sometimes accompanied 
with the inversion 
of the magnetization at the core perpendicular to the disk \cite{Ono}.
Therefore, the dynamics of the magnetic structure induced by the current is 
an important and fundamental issue universal in the metallic magnetic systems.
On the other hand, there are several metallic spiral magnets with the frustrated 
exchange interactions such as Ho metal \cite{Koehler, Cowley}, and with the Dzyaloshinskii-Moriya(DM)
interaction such as MnSi \cite{ MnSi1,MnSi2,MnSi3}, 
(Fe,Co)Si \cite{Uchida}, and FeGe \cite{Uchida2}. 
The quantum disordering under pressure or the nontrivial magnetic textures
have been discussed for the latter class of materials. An important feature is that the 
direction of the wavevector is one of the degrees of freedom in addition to 
the phase of the screw spins. Also the non-collinear nature of the spin 
configuration make it an interesting arena for the study of Berry phase
effect \cite{Berry}, which appears most clearly in the coupling to the 
current. 
However, the studies on the current-induced dynamics of the magnetic
structures with finite wavenumber, e.g., antiferromagnet and spiral magnet,
are rather limited compared with those on the ferromagnetic materials. 
One reason is that the observation of the magnetic DW has been difficult in the 
case of antiferromagnets or spiral magnets. 
Recently, the direct space-time observation of the spiral structure 
by Lorentz microscope becomes possible for the DM induced spiral magnets 
\cite{Uchida,Uchida2} since the wavelength of the spiral is long $( \sim 100 {\rm nm})$.
Therefore, the current-induced dynamics of spiral magnets is now an interesting
problem of experimental relevance.

In this paper, we study the current-induced dynamics of the spiral magnet with the 
DM interaction as an explicit example. 
One may consider that the spiral magnet can be regarded as the periodic
array of the DW's in ferromagnet, but it has many nontrivial 
features unexpected from this naive picture as shown below. 

The Hamiltonian we consider is given by \cite{Landau}
\be
H=\int d{\vec r}\Bigl[{J \over 2}({\vec \nabla} {\vec S})^2+\gamma {\vec S}\cdot ({\vec \nabla} \times {\vec S})\Bigr],
\label{Hcont}
\ee 
where $J>0$ is the exchange coupling constant and $\gamma $ is the 
strength of the DM interaction. 
The ground state of $H$ is realized when ${\vec S}({\vec r})$ 
is a proper screw state such that
\be
{\vec S}({\vec r})=S({\vec n}_1\cos {\vec k}\cdot {\vec r}+{\vec n}_2\sin {\vec k}\cdot {\vec r}),
\ee
where the wavenumber ${\vec k}={\vec n}_3|\gamma |/J$, and ${\vec n}_i (i=1,2,3)$ form the 
orthonormal vector sets. The ground state energy is given by $-VS^2\gamma ^2/2J$ where $V$ is 
the volume of the system. The sign of $\gamma $ is equal to that of 
$({\vec n}_1 \times {\vec n}_2)\cdot {\vec n}_3 $, determining 
the chirality of the spiral.  

The equation of motion of the spin under the current is written as
\be
\dot{\vec S}
={g \mu _{\rm B}\over \hbar}{\vec B}_{\rm eff}\times {\vec S}
-{a^3 \over 2eS}({\vec j}\cdot {\vec \nabla} ){\vec S}
+{a^3 \over 2eS}\beta {\vec S} \times ({\vec j}\cdot {\vec \nabla} ){\vec S}
+{\alpha \over S}{\vec S} \times \dot{\vec S}
\label{LLG}
\ee
where ${\vec B}_{\rm eff}=-\delta H/\delta {\vec S}$ is the effective 
magnetic field and $\alpha $, $\beta$ are the Gilbert damping constants
introduced phenomenologically \cite{Zhang, Thiaville}.

We discretize the Hamiltonian Eq.(1) and the equation of motion Eq.(3) by 
putting spins on the chain or the square lattice with the lattice constant $a$, and
replacing the derivative by the difference. 
The length of the spin $| {\vec S}_i|$ is a constant of motion at each site $i$, and we can easily derive
$
{\dot H}=\frac{\delta H}{\delta {\vec S}}\cdot {\dot {\vec S}}
=-\alpha |{\dot {\vec S}}|^2
$
from Eq.(3), i.e., the energy continues to decrease as the time evolution.

We start with the one-dimensional case along $x$-axis as shown in Fig.1. 
The discretization means replacing $\partial _x{\vec S}(x)$ by 
$({\vec S}_{i+1}-{\vec S}_{i-1})/2a$, and $\partial ^2_x{\vec S}(x)$ by 
$({\vec S}_{i+1}-2{\vec S}_{i}+{\vec S}_{i-1})/a^2$. We note that the wavenumber 
which minimizes Eq.(\ref{Hcont}) is $k=k_0=\arcsin(\gamma /J)$
on the discretized one-dimensional lattice.
Numerical study of Eq.(3) have been done with $g \mu _{\rm B}/ \hbar=1$, $2e=1$, $S=1$, $a=1$
$J=2$, and $\gamma =1.2$. 
In this condition, the wavelength of the spiral 
$\lambda =2\pi /k_0 \approx 11.6$ is long compared with the lattice constant $a=1$, 
and we choose the time scale $\Delta t /(1+\alpha ^2)= 10^{-2}$. 
We have confirmed that the results do not depend on $\Delta t$ even if it is reduced by the factor 
$10^{-1}$ or $10^{-2}$. The sample size $L$ is $10^4$ with the open boundary condition. 
As we will show later, the typical value of the current is $j\sim 2\gamma $ and in the real situation 
with the wavelength $\lambda [{\rm nm}]$, the exchange coupling constant $J [{\rm eV}]$ 
and the lattice constant $a [{\rm nm}]$, it is 
$j \approx 3.2\times 10^{15}J/(\lambda a) [{\rm A/m^2}]$. 
Substituting $J=0.02$, $\lambda =100$, $a=0.5$ into above estimatation, the typical current is 
$10^{12} [{\rm A/m^2}]$, and the unit of the time is
$\Delta t = J/\hbar \approx 30[{\rm ps}]$. 

The Gilbert damping coefficients $\alpha$, $\beta$ are typically $10^{-3} \sim 10^{-1}$ in the 
realistic systems. In most of the calculations, however, we take $\alpha =5.0$ to accelarate the 
convergence to the steady state. The obtained steady state depends only on the ratio $\beta /\alpha$ 
except the spin configurations near the boundaries as confirmed by the simlations with $\alpha =0.1$.
We employ the two types of initial condition, i.e., the ideal proper screw state with the 
wavenumber $k_0$, and the random spin configurations. The difference of the dynamics in 
these two cases are limited only in the early stage ($t<5000\Delta t$).

Now we consider the steady state with the constant velocity 
for the shift of the spiral pattern obtained after 
the time of the order of $10^5\Delta t$. 
One important issue here is the current-dependence of the velocity, which
has been discussed intensively for the DW motion in ferromagnets.
In the latter case, there appears the intrinsic pinning in the case of 
$\beta=0$ \cite{TataraKohno}, 
while the highly nonlinear behavior for $\beta/\alpha \ne 0$ \cite{Thiaville}.
In the special case of $\beta=\alpha$, the trivial solution corresponding to 
the parallel shift of the ground state configuration of Eq.(1) with the 
velocity $v =j$ is considered to be realized \cite{Barnes}.
Figure 2 shows the results for the velocity, the induced uniform magnetization 
$S_x$ along $x$-axis, and the wavevector $k$ of the spiral in the steady state.
The current-dependence of the velocity for the cases of 
$\beta=0.1, 0.5 \alpha$, $\alpha $ and $2\alpha$ is shown in Fig. 2(a).
Figure 2(b) shows the 
$\beta/\alpha$-dependence of the velocity for the fixed current $j=1.2$.
It is seen that the velocity is almost proportional to both the current $j$ and
the ratio $\beta/\alpha$. 
Therefore, we conclude that the velocity $v=(\beta/\alpha )j$ 
without nonlinear behavior up to the current $j\sim 2\gamma $, which is in 
sharp contrast to the case of the DW motion in ferromagnets.
The unit of the velocity is given by $a/\Delta t$, which is of the order of 
$20 [{\rm m/s}]$ for $a \approx 5 [{\rm \AA}]$ and $\Delta t \approx 30 [{\rm ps}]$.
In Fig. 2(c) shown the wavevector $k$ of the spiral under the current $j=1.2$ for 
various values of $\beta /\alpha$. 
It shows a non-monotonous behavior with the maximum at $\beta /\alpha \approx 0.2$, and 
is always smaller than the wavenumber $k_0$ in the equilibrium shown in the dotted line.
Namely, the period of the spiral is elongated by the current. 
As shown in Fig. 2(d), there appears the uniform magnetization $S_x$
along the $x$-direction. $S_x$ is zero and changes the sign at $\beta /\alpha =1$. 
With the positive $\gamma$ (as in the case of Fig. 2(d)), $S_x$ is anti-parallel to 
the current $j//x$ with $\beta < \alpha$ and changes its direction for $\beta > \alpha$. 
For the negative $\gamma$, the sign of $S_x$ is reversed. As for the velocity ${\vec v}$, 
on the other hand, it is always parallel to the current ${\vec j}$.

\begin{figure}
\includegraphics[width=5.1cm]{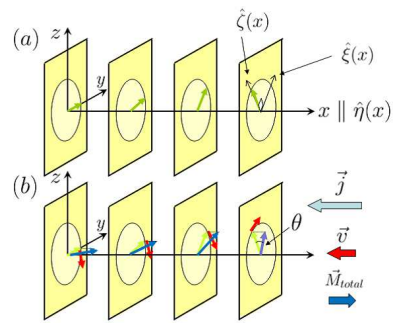}
\caption{Spin configurations in the spiral magnet (a) in the equilibrium state, 
and (b) under the current. Under the current ${\vec j}$, the uniform magnetization $S_x$ along 
the spiral axis/current direction is induced together with the rotation of the
spin, i.e., the parallel shift of the spiral pattern with the velocity $v$.
Note that the magnetization is anti-parallel/parallel to the current 
direction with positive/negative $\gamma $ for $\beta < \alpha$, while it is reversed
for $\beta > \alpha $, and the wavenumber $k$ changes
from the equilibrium value. }
\label{screw_spiral}
\end{figure}

\begin{figure}
\begin{tabular}{cc}
\includegraphics[width=4.2cm]{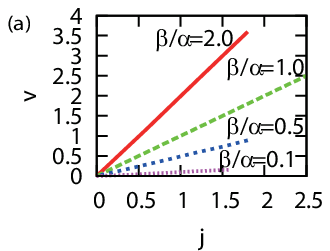} & \hspace{1em}
\includegraphics[width=3.8cm]{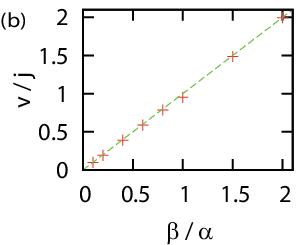} \hspace{1em} \\
\includegraphics[width=3.8cm]{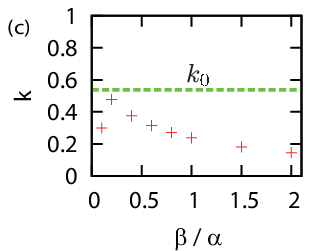} & \hspace{1em}
\includegraphics[width=3.8cm]{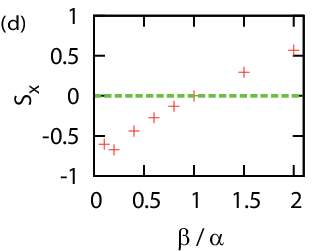} \hspace{1em}
\end{tabular} 
\caption{For the case of $\gamma =1.2$, the numerical result is shown. 
(a) The steady state velocity $v$ as a function of the current $j$ with 
the fixed values of $\beta=0.1, 0.5\alpha$, $\alpha$, and $2\alpha$.   
(b) The velocity $v$ as a function of $\beta/\alpha$ for a fixed value
of the current $j=1.2$. $v$ is almost proportional to $\beta/\alpha$. 
(c) The wavenumber $k$ as a function of $\beta/\alpha$. The dotted line shows
$k_0$ in the equilibrium.
(d) The uniform magnetization $S_x$ along the current direction as a 
function of $\beta/\alpha$ for a fixed value of $j=1.2$.
}
\label{beta}
\end{figure}

Now we present the analysis of the above results in terms of the continuum theory and a scaling argument.
For one-dimensional case, the modified LLG Eq.(3) can be recast in the following form:
\be
{\dot {\vec S}}=-J{\vec S}\times\partial^2_x{\vec S}-(2\gamma S_x+j)\partial_x{\vec S}+{\vec S}\times(\alpha{\dot {\vec S}}+\beta j \partial_x {\vec S}).
\label{1DLLG}
\ee
It is convenient to introduce a moving coordinates ${\hat \xi}(x,t)$, ${\hat \eta}$, and ${\hat \zeta}(x,t)$ (see Fig.\ref{screw_spiral}) \cite{Nagamiya}. They are explicitly defined through ${\hat x}$, ${\hat y}$ and ${\hat z}$ as
\begin{eqnarray}
{\hat \zeta}(x,t)&=&\cos(k(x-vt)+\phi){\hat y}+\sin(k(x-vt)+\phi){\hat z}, \nonumber \\
{\hat \xi}(x,t)&=&-\sin(k(x-vt)+\phi){\hat y}+\cos(k(x-vt)+\phi){\hat z}, \nonumber
\end{eqnarray} 
and ${\hat \eta}={\hat x}$. We restrict ourselves to the following ansatz:
\be
{\vec S}(x,t)=S_x {\hat \eta} + \sqrt{1-S^2_x}{\hat \zeta}(x,t),
\label{1Dansatz}
\ee
where $S_x$ is assumed to be constant. 

By substituting Eq.(\ref{1Dansatz}) into 
Eq.(\ref{1DLLG}), we obtain $v$ as 
\be
v = \frac{\beta}{\alpha}j,
\label{v} 
\ee
by requiring that there is no force along ${\hat \eta}$- and ${\hat \zeta}$-directions 
acting on each spin.
In contrast to the DW motion in ferromagnets, the velocity $v$ becomes zero when 
$\beta \to 0$ even for large value of the current. The numerical results in 
Fig. 2(a), (b) show good agreement with this prediction Eq.(6). 

On the other hand, the magnetization $S_x$ along $x$-axis is given by 
\be
S_x = { { \beta/\alpha -1} \over {2\gamma -J k}}j, 
\label{Sx} 
\ee
once the wavevector $k$ is known. 
Here we note that the above solution is degenerate with respect to $k$, which needs 
to be determined by the numerical solution. 
From the dimensional analysis, the spiral wavenumber $k$ is given by the scaling form, 
$k=k_0 g(j/(2\gamma),\beta/\alpha)$ with the dimensionless function $g(x,y)$ and 
also is $S_x$ through Eq.(7).

Motivated by the analysis above, we study the $\gamma $-dependence of 
the steady state properties. 
In Fig.3, we show the numerical results for $k/k_0$ and $S_x$ as the functions of $j/2\gamma$ 
in the cases of $\beta / \alpha =0.1, 0.5$ and $2$. 
Roughly speaking, the degeneracies of the data are obtained approximately for each color points
(the same $\beta /\alpha$ value) with different $\gamma$ values.
The deviation from the scaling behavior is due to the discrete nature of the lattice model, which
is relevant to the realistic situation. 
For $\beta / \alpha =0.1$ (black points in Fig.3), $k$ remains constant and $S_x$ is induced almost
proportional to the current up $j/2\gamma \approx 0.4$, where the abrupt change of $k$ occurs. 
For $\beta / \alpha =0.5$ (blue points) and $\beta / \alpha =2.0$ (red points), the changes of $k$
and $S_x$ are more smooth. A remarkable result is that the spin $S_i$ on the lattice point $i$ is well
described by Eq.(5) at $x=x_i$, and hence the relation Eq.(7) is well satisfied as shown by the
curves in Fig. 3(b), even though the scaling relation is violated to some extent. For larger values 
of $j$ beyond the data points, i.e., $j/2\gamma >0.75$ for $\beta /\alpha =0.1$, $j/2\gamma >1.5$ 
for $\beta /\alpha =0.5$ and $j/2\gamma > 0.9$ for $\beta /\alpha =2.0$, the spin configuration is 
disordered from harmonic spiral characterized by a single wavenumber $k$. The spins are the chaotic
funtion of both space and time in this state analogous to the turbulance. This instability is 
triggered by the saturated spin $S_x=\pm 1$, occuring near the edge of the sample. 

\begin{figure}
\includegraphics[width=8cm]{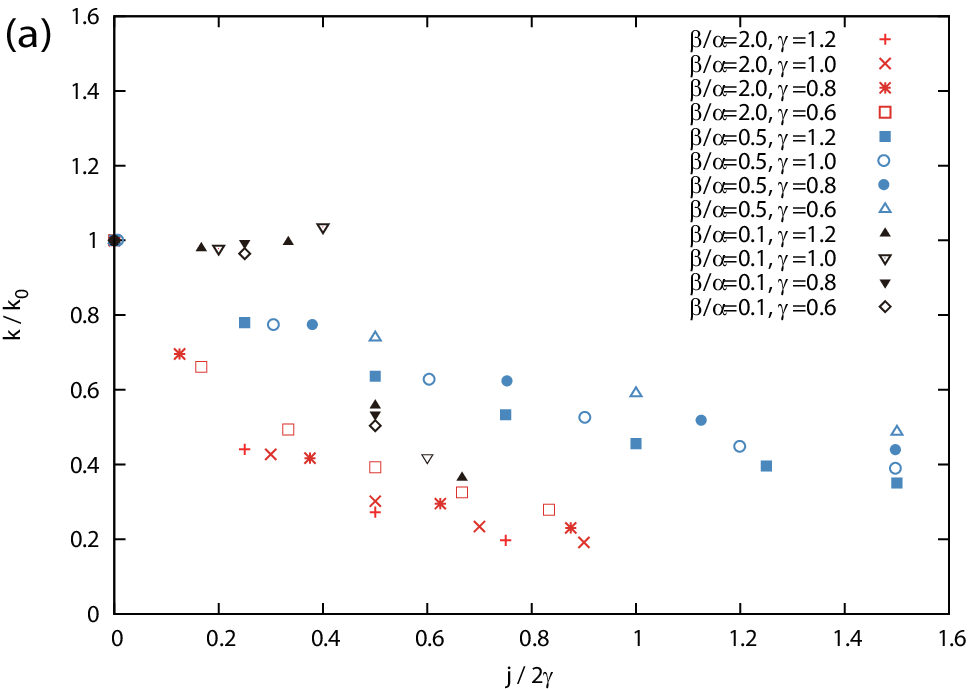}
\includegraphics[width=8cm]{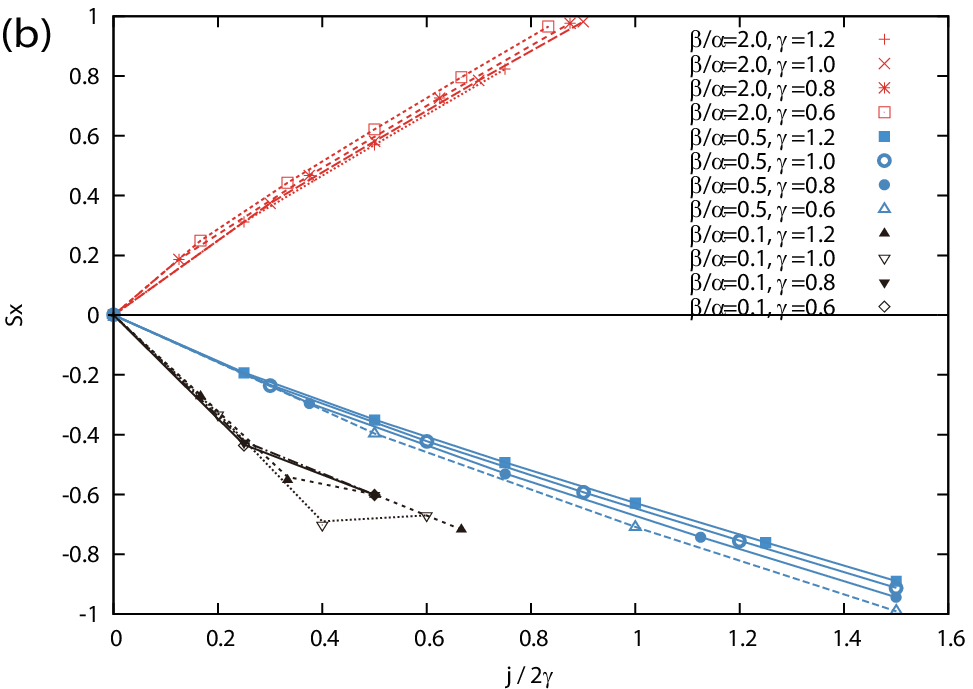}
\caption{The scaling plot for (a) $k/k_0$ where $k_0$ is the wavenumber in the equilibrium without
the current, 
and (b) $S_x$ in the steady state as the function of $j/2\gamma $. The black, blue, and 
red color points correspond to $\beta /\alpha =0.1$, 
$0.5$ and $2.0$, respectively. The curves in (b) indicate Eq.(7) calculated from the $k$ values 
in (a), showing the good agreement with the data points.}
\label{k}
\end{figure}

\begin{figure}
(a) ${\vec j}=0$
\\
\includegraphics[width=8cm]{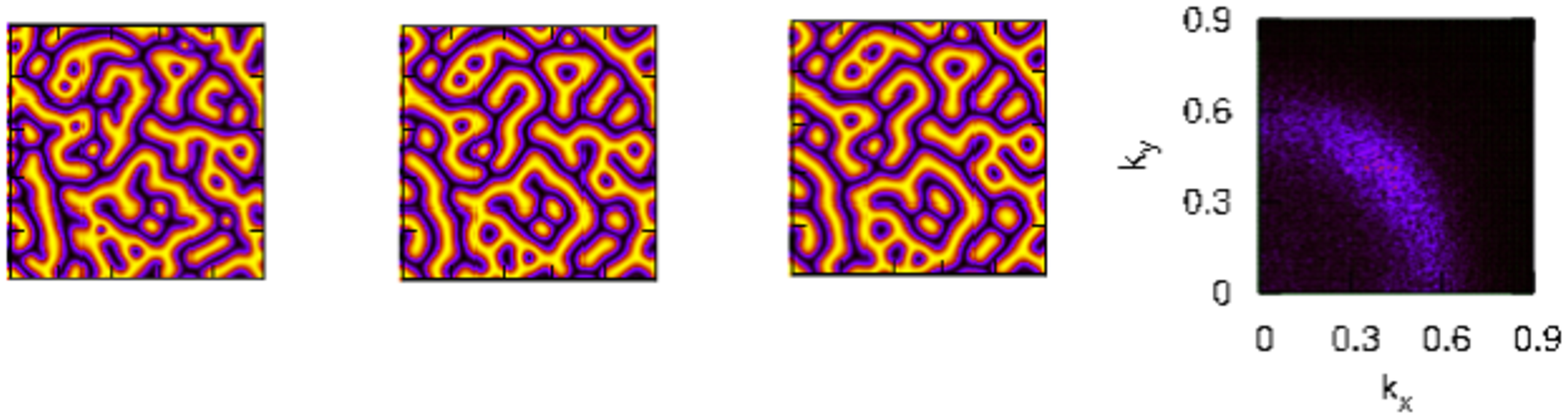}
\\
(b) ${\vec j}=(0.3, 0)$
\\
\includegraphics[width=8cm]{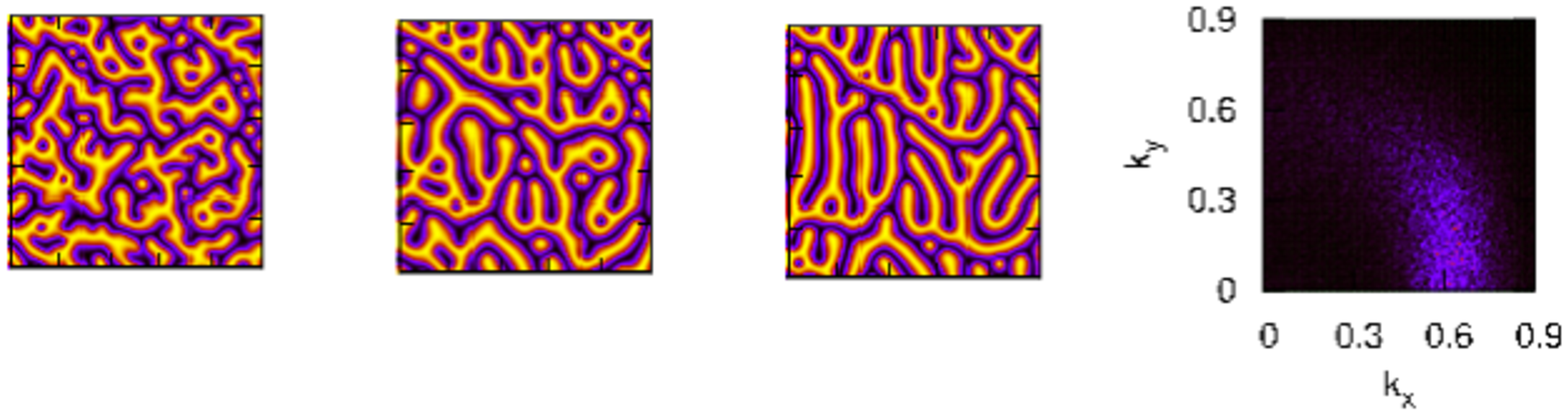}
\\
(c) ${\vec j}=(0.3/\sqrt{2}, 0.3/\sqrt{2})$
\\
\includegraphics[width=8cm]{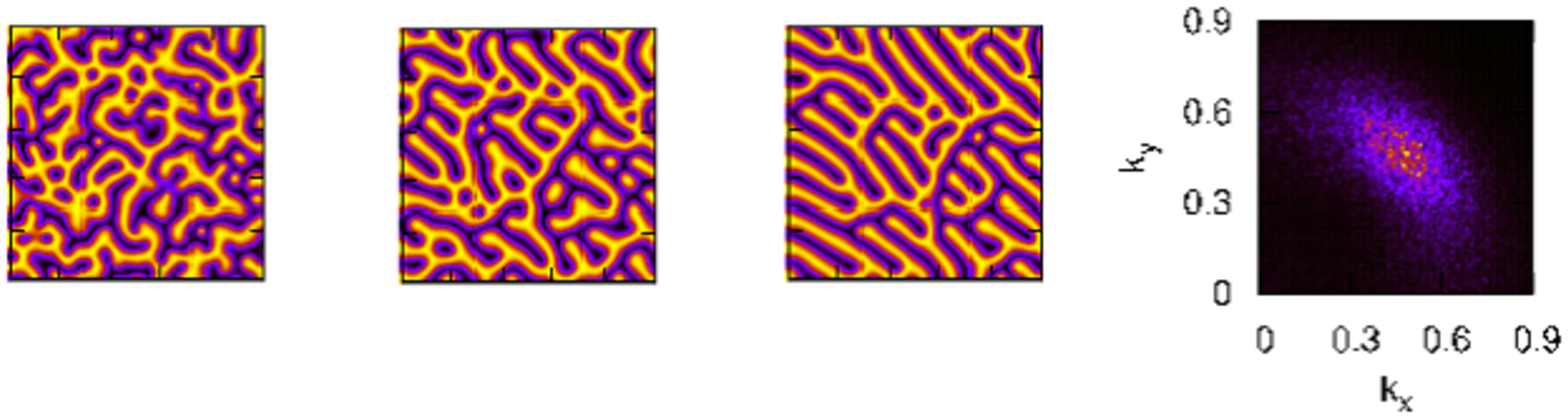}
\\
color box of $S_z({\bf r})$
\includegraphics[width=7.4cm]{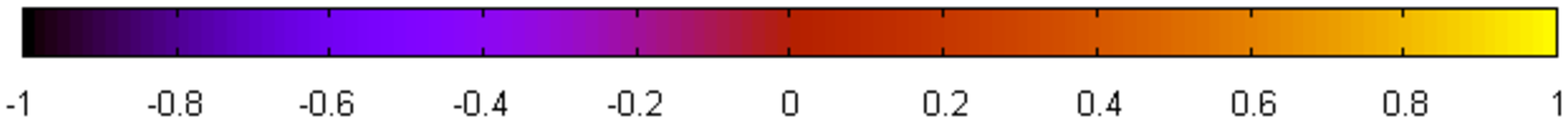}
\\
color box of $|S_z({\bf k})|^2$
\includegraphics[width=8cm]{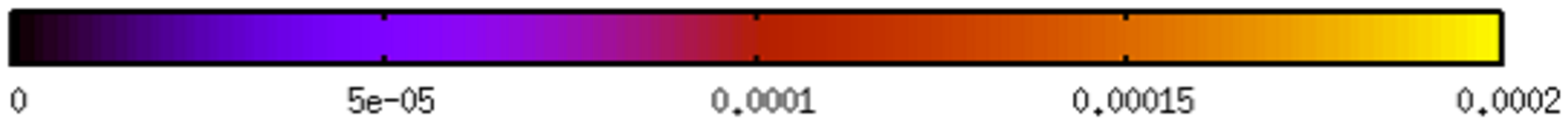}
\caption{The time evolution of the $z$ component $S_z$ of the spin from the random initial 
configuration of the $10^2 \times 10^2$ section in the middle of the sample is shown in the 
case of (a) $j=0$, (b) $j=0.3$ along the $x$-axis, (c) $j=0.3$ along the (1,1)-direction. 
From the left, $t=10^2\Delta t$, $1.7\times 10^3\Delta t$, $5\times 10^3\Delta t$. 
The rightmost panels show the spectral intensity $|S_z({\vec k},5\times 10^3\Delta t)|^2$ 
from the whole sample of the size $2^{10} \times 2^{10}$ in the momentum space ${\vec k}=(k_x, k_y)$. }
\label{2d-eps}
\end{figure}

Next, we turn to the simulations on the two-dimensional square lattice 
in the $xy$-plane. In this case, the direction of the spiral wavevector becomes
another important variable because the degeneracy of the ground state energy occurs. 

Starting with the random spin configuration, we simulate the 
time evolution of the system without and with the current as shown in 
Fig.4. Calculation has been done with the same parameters as in the one-dimensional case 
where $\gamma =1.2$, $\beta=0$, and the system size is $2^{10}\times 2^{10}$.

In the absence of the current, the relaxation of the spins into the 
spiral state is very slow, and many dislocations remain even after a 
long time. Correspondingly, the energy does not decrease to the 
ground state value but approaches to the higher value with the 
power-law like long-time tail. The momentum-resolved intensity 
is circularly distributed with the broad width as shown in Fig. 4(a) 
corresponding to the disordered direction of ${\vec k}$.
This glassy behavior is distinct from the relaxation dynamics of the 
ferromagnet where the large domain formation occurs even though the 
DW's remain.  
Now we put the current along the  ${\hat x}$ (Fig. 4(b)) and 
 $({\hat x}+{\hat y})$ (Fig. 4(c)) directions. 
It is seen that the direction of ${\vec k}$ is controlled by the current
also with the radial distribution in the momentum space being narrower
than that in the absence of $j$ (Fig. 4(a)). This result suggests that the 
current $j$ with the density $\sim 10^{12} [{\rm A/ m^2}]$ of the time duration 
$\sim 0.1 [\mu {\rm sec}]$ can anneal the directional disorder of the spiral magnet. 
After the alignment of ${\vec k}$ is achieved, the simulations on the one-dimensional 
model described above are relevant to the long-time behavior.

To summarize, we have studied the dynamics of the 
spiral magnet with DM interaction under the current $j$ by solving the 
Landau-Lifshitz-Gilbert equation numerically. 
In the steady state under the charge current $j$, the velocity $v$ is given by $(\beta/\alpha)j$ 
($\alpha, \beta$: the Gilbert-damping coefficients), the 
uniform magnetization is induced parallel or anti-parallel to the current direction, 
and period of the spiral is elongated.  
The annealing effect especially on the direction of the spiral 
wavevector ${\vec k}$ is also demonstrated.
 
The authors are grateful to N. Furukawa and Y. Tokura for fruitful discussions.
This work was supported in part by Grant-in-Aids (Grant No. 15104006, No. 16076205, and No. 17105002) and NAREGI Nanoscience Project from the Ministry of Education, Culture, Sports, Science, and Technology.
HK was supported by the Japan Society for the Promotion of Science.

\end{document}